\documentclass[twoside]{aiml}

\usepackage{aimlmacro}

\usepackage{amssymb}
\usepackage{amsmath}
\usepackage{mathtools}
\usepackage[inline]{enumitem}
\usepackage[colorlinks=true]{hyperref}
\usepackage{tikz}
\usetikzlibrary{positioning,calc,arrows,quantikz}
\usepackage{tikz-cd}

\usepackage{calc}
\usepackage{longtable}

\usepackage{scalerel}

\newcommand{\bbsemicolon}{%
  \scalerel*{%
    \hbox{\usefont{U}{bbold}{m}{n} ;}%
  }{;}%
}
\newcommand{\comp}{\mathbin{\bbsemicolon}}

\usepackage{array,etoolbox}
\newcounter{nrf}

\makeatletter

\newlength{\PS@lastparam}
\newlength{\PSlastparam}
\newcommand{\PSlp}{%
  \setlength{\PSlastparam}{\PS@lastparam}%
  \the\PSlastparam
}
\def\PS@sub@lastparam{}

\newcommand{\PS@numwidth}{99}
\newcommand{\PSnumwidth}[1]{%
  \renewcommand{\PS@numwidth}{#1}%
}

\newcommand{\PS@style}{\small}
\newcommand{\PS@numstyle}{\footnotesize}

\newlength{\PSindent}
\newlength{\PS@extraindent}
\newlength{\PSpre}
\newlength{\PSpost}
\newlength{\PS@Nwidth}
\newlength{\PS@Swidth}
\newlength{\PS@Ewidth}
\newlength{\PScolsep}
\newcommand{\PS@rownumber}{%
  \ifPS@subsubsteps
  \thePSsubstepc.%
  \the\numexpr\value{PSsubsubstepc}+1\relax
  \else
  \ifPS@substeps
  \thePSstepc.%
  \the\numexpr\value{PSsubstepc}+1\relax
  \else
  \the\numexpr\value{PSstepc}+1\relax
  \fi\fi
}
\newcommand{\PS@step}{%
  \ifPS@subsubsteps
  \refstepcounter{PSsubsubstepc}%
  \else
  \ifPS@substeps
  \refstepcounter{PSsubstepc}%
  \else
  \refstepcounter{PSstepc}
  \fi\fi%
}

\newif\ifPS@inprogress
\newif\ifPS@substeps
\newif\ifPS@subsubsteps
\newif\ifPS@continued
\newif\ifPS@subcontinued
\newcounter{PSc}
\newcounter{PSstepc}[PSc]
\newcounter{PSsubstepc}[PSstepc]
\renewcommand{\thePSsubstepc}{\thePSstepc.\arabic{PSsubstepc}}
\newcounter{PSsubsubstepc}[PSsubstepc]

\newenvironment{proofsteps}[1]{%
  \global\settowidth{\PS@lastparam}{\PS@style\hspace*{#1}}
  \ifPS@continued\else\refstepcounter{PSc}\fi
  \begingroup
  \setlength{\LTpre}{\PSpre}%
  \setlength{\LTpost}{\PSpost}%
  
  \setlength{\tabcolsep}{0pt}
  \noindent\PS@style
  \settowidth{\PS@Nwidth}{\PS@numstyle\PS@numwidth}%
  \setlength{\PS@Swidth}{#1}%
  \addtolength{\PS@Swidth}{-\PS@extraindent}%
  \setlength{\PS@Ewidth}{\linewidth}%
  \addtolength{\PS@Ewidth}{-\PSindent}%
  \addtolength{\PS@Ewidth}{-\PS@extraindent}%
  \addtolength{\PS@Ewidth}{-\PS@Nwidth}%
  \addtolength{\PS@Ewidth}{-\PScolsep}%
  \addtolength{\PS@Ewidth}{-\PS@Swidth}%
  \addtolength{\PS@Ewidth}{-\PScolsep}%
  \PS@inprogresstrue
  \longtable{%
    @{\hspace*{\PSindent}\hspace*{\PS@extraindent}\makebox[\PS@Nwidth][r]{\PS@rownumber}}%
    @{\hskip\PScolsep}>{\PS@step}p{\PS@Swidth}%
    @{\hskip\PScolsep}>{\footnotesize\raggedright\arraybackslash}p{\PS@Ewidth}%
  }%
}{%
  \ifPS@inprogress
  \addtocounter{table}{-1}%
  \endlongtable  
  \endgroup
  \PS@continuedfalse
  \PS@inprogressfalse
  \else\fi
}

\newcommand{\PSbreak}[1]{%
  \endproofsteps
  \par\medskip
  #1
  \medskip\par
  \PS@continuedtrue
  \proofsteps{\PS@lastparam}%
}

\newif\ifPS@sub@inprogress

\newif\ifPS@laststep
\newcommand{\laststep}{\global\PS@laststeptrue}
\newif\ifPS@lastsubstep
\newcommand{\lastsubstep}{\global\PS@lastsubsteptrue}

\newcommand{\adjustcol}[1]{%
  \global\advance\@colroom-#1%
}

\makeatother

\newcommand*{\pcformat}[1]{%
  [\;{\normalfont\itshape #1}\;]%
}

\newenvironment{proofcases}[1][]{%
  \description[font=\pcformat, leftmargin=\parindent, #1]%
}{\enddescription}

\newcommand{\PL}{{\mathsf{PL}}}

\newcommand{\HDQL}{{\mathsf{HDQL}}}
\newcommand{\Sig}{\mathtt{Sig}}
\newcommand{\Mod}{\mathtt{Mod}}
\newcommand{\Sen}{\mathtt{Sen}}
\newcommand{\Set}{\mathbb{S}et}
\newcommand{\Cat}{\mathbb{C}at}

\newcommand{\Prop}{\mathtt{Prop}}

\newcommand{\C}{{\mathbb{C}}}
\newcommand{\N}{\mathbb{N}}
\newcommand{\W}{\mathcal{W}}
\newcommand{\X}{\mathcal{X}}
\newcommand{\Y}{\mathcal{Y}}
\newcommand{\Hil}{\mathcal{H}}

\newcommand{\act}{\mathfrak{a}}
\newcommand{\bact}{\mathfrak{b}}

\newcommand{\cn}{\mathtt{c}}
\newcommand{\hil}{\mathtt{h}}
\newcommand{\vc}{\mathtt{v}}
\newcommand{\ut}{\mathtt{u}}
\newcommand{\w}{\mathtt{w}}

\newcommand{\red}{\!\upharpoonright\!}
\newcommand{\at}[1]{@_{#1}\,}
\newcommand{\nec}[1]{[#1]}
\newcommand{\pos}[1]{\langle #1 \rangle}
\newcommand{\store}[1]{{\downarrow}#1\,{\cdot}\,}

\newcommand{\ip}[2]{\langle #1  \mid #2\rangle}
\newcommand{\Frac}[2]{\displaystyle\frac{#1}{#2}}
\newcommand{\Space}{\hspace{0.5cm}}

\def\lastname{D. G\u{a}in\u{a}}

\begin{document}

\begin{frontmatter}
\title{Birkhoff style proof systems for hybrid-dynamic quantum logic}
\author{Daniel G\u{a}in\u{a}}\footnote{This work has been partially supported by Japan Society for the Promotion of Science, grant number 23K11048.}
 \address{Kyushu University}
 
\begin{abstract}
We explore a simple approach to quantum logic based on hybrid and dynamic modal logic, where the set of states is given by some Hilbert space.
In this setting, a notion of quantum clause is proposed in a similar way the notion of Horn clause is advanced in first-order logic, that is, to give logical properties for use in logic programming and formal specification.
We propose proof rules for reasoning about quantum clauses and we investigate soundness and compactness properties that correspond to this proof calculus.
Then we prove a Birkhoff completeness result for the fragment of hybrid-dynamic quantum logic determined by quantum clauses.
\end{abstract}

\begin{keyword}
quantum clause, Birkhoff completeness, hybrid-dynamic logic, algebraic specification, institution
\end{keyword}
\end{frontmatter}

\section{Introduction}
The logical framework for studying quantum theory was originally proposed by Birkhoff and von Neumann~\cite{birk-neu36}. 
The truth values of this quantum logic are given by the elements of a Hilbert lattice, which is the lattice consisting of all closed subspaces of a Hilbert space.
A more modern view on the semantics of quantum logic is based on Kripke structures and modal logics.
One can express in a modal logic (based on a local Boolean satisfaction) quantum properties captured traditionally by (non-Boolean) Quantum Logic.
For example, the \emph{orthocomplement}, also called \emph{quantum negation},  $\sim\varphi$, is defined as the set of all vectors orthogonal on the vectors where $\varphi$ holds, while \emph{quantum disjunction} $\varphi_1\oplus \varphi_2$ is defined by $\sim(\sim\varphi_1\wedge \sim\varphi_2)$.
There are many approaches based on extensions of:
\begin{enumerate}
\item dynamic logic such as  Dynamic Quantum Logic~\cite{bru-jor2004}, Logic of Quantum Actions~\cite{BaltagS2005} and Logic of Quantum Programs~\cite{DBLP:journals/mscs/BaltagS06},
\item temporal logic such as Quantum Linear Temporal Logic~\cite{mateus-2009} and Quantum Computation Tree Logic~\cite{balt2008}, and
\item both dynamic and temporal logic such as Linear Temporal Quantum Logic proposed in~\cite{takagi2023}. 
\end{enumerate}
In this paper, we study a variant of quantum logic with features from both hybrid and dynamic logics which was originally proposed in~\cite{gai-qinit}.

Dynamic propositional logic is suitable for reasoning about classical programs.
Naturally, quantum versions of dynamic propositional logic were developed to reason about quantum programs.
On the other hand, hybrid logics are known for their ability to name individual states of Kripke structures, which allow a more uniform proof theory~\cite{brau11,DBLP:journals/jacm/Gaina20} and model theory~\cite{DBLP:journals/apal/GainaBK23,DBLP:journals/tcs/Gaina17,GainaBKG22} than non-hybrid modal logics.
Hybrid logics are equipped with features to distinguish states and reason about their properties, which in turn is important to applications in formal methods.
In addition, one can express temporal properties using sentence operators inherent to hybrid logics such as store and retrieve~\cite{DBLP:journals/jsyml/ArecesBM01}.
Therefore, the variant of hybrid-dynamic quantum logic studied in this paper allows one to express both temporal and dynamical properties.

The users of a quantum programming language must describe the dynamics of quantum systems by relying on data types consisting of scalars and vectors of a concrete Hilbert space.
Therefore, a necessary feature of a quantum programming language is the existence of some predefined data types for the scalars and vectors used in applications.
From a model theoretic point of view, Hilbert spaces are vectorial spaces equipped with an inner product such that each Cauchy sequence of vectors has a limit.
Since this is not a first-order property, the definition of Kripke structures whose states are given by some Hilbert space and the development of the logical results for them are significantly more difficult. 
Following the ideas advanced in \cite{gai-qinit},
we employ the method of diagrams proposed by Robinson in model theory to define Hilbert spaces and Kripke structures over them.
We use constant symbols to stand for the elements of the Hilbert space to be constructed, and we work within the theory which contains all the equations and relations satisfied by that Hilbert space.
This means that the signature of nominals used to describe frames in hybrid logics is replaced by positive diagrams of concrete Hilbert spaces.
The diagrams of Hilbert spaces can be regarded as the counterparts of libraries defining scalars and vectors from quantum programming, since individual elements of Hilbert spaces can be named in the hybrid-dynamic quantum logic studied in this paper.
This is an important feature which brings the present work closer to applications in formal methods.

In this paper, we provide sound and complete proof calculi for a fragment of hybrid-dynamic quantum logic of \cite{gai-qinit}. 
The sentences are restricted to  \emph{quantum clauses}, which are obtained from propositional symbols by applying the following sentence operators:
both quantum and classical implication,
necessity over structural \emph{actions} which, in turn, are constructed from \emph{projective measurements} and \emph{unitary transformations}, 
etc.
In addition, we provide proof rules to reason formally about the properties of those Kripke structures that are specified using quantum clauses.
To conclude, the main result of the paper is a completeness theorem for the fragment of hybrid-dynamic quantum logic obtained by restricting the sentences to quantum clauses.

A brief comparison with the work recently reported in~\cite{gai-qinit} is also in order: 
both papers deal with properties of hybrid-dynamic quantum logic 
(however, \cite{gai-qinit} is the contribution in which we introduced the logic);
and in both papers we examine quantum clauses;
but the results that we develop are complementary: 
in~\cite{gai-qinit}, we focused on an initiality result and on Herbrand's theorem, whereas here we advance proof calculi for the logic.
This latter endeavour is much more complex, because it deals with syntactic entailment instead of semantic entailment.
In \cite{DBLP:journals/fac/Gaina17}, the author proves a Birkhoff completeness result in the abstract framework given by notion of stratified institution~\cite{aig-strat,dia-ult-kripke}, which is a category-based formalization of the notion of modal logic.
Another completeness result for Horn clauses can be found in \cite{DBLP:conf/tableaux/GainaT19} for a variant of hybrid-dynamic first-order logic with user-defined sharing.
It is worth noting that unlike \cite{DBLP:journals/fac/Gaina17} and \cite{DBLP:conf/tableaux/GainaT19}, in this work, cut rule is not used to prove completeness, which means that lemma discovery is not needed in formal proofs.

According to \cite{Baltag2011-BALQLA}, the correct semantics of quantum-logical connectives is in terms of dynamic modalities, rather than purely propositional operators.
This philosophical interpretation is supported by some technical developments reported in \cite{BaltagS2005,DBLP:journals/mscs/BaltagS06,DBLP:journals/sLogica/BaltagS08}.
The present work is based on the same ideas, but it departs from any of those studies due to the fact that the set of states is not the set of one-dimensional closed subspaces of some Hilbert space but the entire set of vectors of a Hilbert space.
The present approach narrows the gap between theory and its applications (e.g., to formal methods), since it allows one to name concrete vectors and scalars and use them to build sentences.

\section{Hilbert spaces} \label{sec:hilbert}
This section is dedicated to a brief presentation of Hilbert spaces from a first-order logic perspective.
The signature of Hilbert spaces $\Sigma^\hil=(S^\hil,F^\hil,P^\hil)$ is a first-order signature that consists of:
\begin{enumerate}
\item a set of sorts $S^\hil=\{\cn,\vc\}$, where $\cn$ denotes the sort of complex numbers and 
$\vc$ denotes the sorts of vectors.

\item a set of function symbols $F^\hil=F^\cn\cup F^\vc$, where 
\begin{enumerate}
\item $F^\cn$ is a set of the usual operations on complex numbers such as 
addition $\_+\_:\cn~\cn\to\cn$,
multiplication $\_*\_:\cn~\cn\to\cn$, etc;
\item $F^\vc=\{ 
\_+\_:\vc~\vc\to\vc, 
0:\to \vc, 
\_\_ : \cn~\vc\to \vc, 
\ip{\_}{\_}:\vc~\vc\to\cn \}$, where 
$\_+\_:\vc~\vc\to\vc$ denotes the vector addition,
$0$ denotes the origin vector, 
the juxtaposition $\_\_ : \cn~\vc\to \vc$ denotes the scalar multiplication, and
$\ip{\_}{\_}:\vc~\vc\to\cn$ denotes the inner product;
\end{enumerate}
\item a singleton $P=\{\_<\_:\cn~\cn\}$ consisting of one relation symbol $\_<\_:\cn~\cn$ denoting the ordering among real numbers.
\end{enumerate}
A Hilbert space is a first-order model $\Hil$ defined over $\Sigma^\hil$ such that:
\begin{itemize}
\item $\Hil_\cn$ is the set of complex numbers $\C$;
the model $\Hil$ interprets all function symbols in $F^\cn$ as the usual operations on complex numbers;

\item $\Hil_\vc$ is a set of vectors,
$+^\Hil:\Hil_\vc\times \Hil_\vc\to \Hil_\vc$ is the vector addition, 
${\_\_}^\Hil: \Hil_\cn\times \Hil_\vc\to \Hil_\vc$ is the scalar multiplication,
$\ip{\_}{\_}^\Hil:\Hil_\vc\times \Hil_\vc\to \Hil_\cn$ is the inner product in which each Cauchy sequence of vectors has a limit.
\footnote{Recall that a sequence of vectors $\{w_n\}_{n\in \N}$ is a Cauchy sequence if for any $\varepsilon >0$  there exists $n\in\N$ such that $||w_i-w_j||<\varepsilon$ for all $i,j\geq n$, where $||w||=\sqrt{\ip{w}{w}}$ is the length/norm of a vector $w\in \Hil_\vc$.}
\item $<^\Hil$ is the usual strict ordering on real numbers.
\end{itemize}
Two vectors $w_1,w_2\in \Hil_\vc$ are orthogonal, in symbols, $w_1\perp w_2$ if $\ip{w_1}{w_2}^\Hil=0^\Hil$.
This notational convention can be extended to sets of vectors in the usual way.
A closed subspace $\X$ of a Hilbert space $\Hil$ is a substructure $\X\subseteq \Hil$ such that
\begin{enumerate*}[label=(\alph*)]
\item $\X_\cn=\C$ and 
\item each Cauchy sequence of vectors  from $\X$ has a limit in $\X$. 
\end{enumerate*}
In other words, a closed subspace is in particular a Hilbert space.
In this paper, we identify a closed subspace by its set of vectors.
The orthocomplement of $\X$ denoted $\X^\perp$ is defined by $\X^\perp=\{w\in \Hil_\vc \mid w\perp \X \}$.
The orthocomplement of a set of vectors is a closed subspace. 
The direct sum of two closed subspaces $\X$ and $\Y$ is defined by $\X\oplus \Y\coloneqq \{x +^\Hil y \mid x\in \X \text{ and } y\in\Y \}$.
The following results are well-known.
\begin{theorem}
Let $\Hil$ be a Hilbert space.
\begin{enumerate}
\item A substructure $\X\subseteq \Hil$ is a closed subspace iff $\X^{\perp\perp}=\X$.
\item $\X\oplus \Y= (\X^\perp \cap \Y^\perp)^\perp$ for all closed subspaces $\X\subseteq \Hil$ and $\Y\subseteq \Hil$.
\item $\X^{\perp\perp}=\Y\cap (\Y\cap \X^\perp)^\perp$ for all closed subspaces  $\X\subseteq \Y\subseteq \Hil$, which means that the (global) closure of $\X$ is equal to local closure of $\X$ relative to $\Y$.
\item $\Hil=\X\oplus \X^\perp$ for all closed subspaces $\X\subseteq \Hil$.
\end{enumerate}
\end{theorem}
By the fourth statement of the theorem above, for any closed subspace $\X\subseteq \Hil$ and any vector $w\in \Hil$, we have $w=w_1 +^\Hil w_2$, where $w_1\in \X$ and $w_2\in \X^\perp$.
Since $\X\cap \X^\perp=\{0^\Hil\}$, the vectors $w_1\in \X$ and $w_2\in \X^\perp$ are uniquely determined.
The projection $P_\X:\Hil\to \X$ is defined by $P_\X(w)=w_1$, where $w=w_1+w_2$, $w_1\in \X$ and $w_2\in \X^\perp$.

\section{Hybrid-Dynamic Quantum Logic} \label{sec:HDQL}
This section is dedicated to the presentation of  Hybrid Dynamic Quantum Logic ($\HDQL$) proposed in \cite{gai-qinit}. 
This logic is an extension of Hybrid-Dynamic Propositional Logic with some constraints on the possible world semantics.

\subsection{Signatures}
The signatures of $\HDQL$ are of the form $\Delta=(\Sigma,E,\Prop)$, where:
\begin{enumerate}
\item $\Sigma=(S^\hil,F^\hil\cup U\cup Q\cup D\cup C,P^\hil)$ is a first-order signature obtained from the signature of Hilbert spaces $\Sigma^\hil=(S^\hil,F^\hil,P^\hil)$ from Section~\ref{sec:hilbert} by adding:
\begin{enumerate}
\item a set $U$ of unitary transformation symbols of the form $u:\vc\to\vc$;
\item a set $Q$ of projective measurement symbols of the form $q:\vc\to\vc$;
\item a set $D$ of constants of sort vector $\vc$; and
\item a set $C$ of constants of sort scalar $\cn$;
\end{enumerate}

\item $E$ is a set of first-order sentences over $\Sigma$;

\item $\Prop$ is a set of propositional symbols which contains a subset $\Prop_c$ of \emph{closed propositional symbols}.
\end{enumerate}
An example of $(\Sigma,E)$ is the positive Robinson diagram of some Hilbert space, that is, all equations and relations satisfied by some Hilbert space.
See Section~\ref{sec:L} for more details.
We make the following notational conventions:
\begin{itemize}
\item We let $\Delta$ range over signatures of the form $(\Sigma,E,\Prop)$ as described above.
\item Similarly, we let $\Delta_i$ range over signatures of the form $(\Sigma_i,E_i,\Prop_i)$, where $\Sigma_i=(S^\hil,F^\hil\cup U_i\cup Q_i\cup D_i \cup C_i,P^\hil)$ and $i$ is any index.
\end{itemize}
Signature morphisms $\chi:\Delta_1\to \Delta_2$ consists of:
\begin{enumerate}
\item a first-order theory morphism $\chi:(\Sigma_1,E_1)\to (\Sigma_2,E_2)$, which is the identity on $\Sigma^\hil$, $\chi(U_1)\subseteq U_2$ and $\chi(Q_1)\subseteq Q_2$,\footnote{This means that $\chi:\Sigma_1\to\Sigma_2$ is a first-order signature morphism such that $E_2$ satisfies $\chi(E_1)$, in symbols,  $E_2\models\chi(E_1)$.} 
and
\item a mapping $\chi:\Prop_1\to\Prop_2$ on propositional symbols, which preserves closed propositional symbols, that is, $\chi(\Prop_1^c)\subseteq \Prop_2^c$.
\end{enumerate}
We overloaded the notation such that $\chi$ denotes not only the signature morphism $\chi:\Delta_1\to \Delta_2$ but also its restrictions $\chi:\Sigma_1\to\Sigma_2$ and $\chi:\Prop_1\to\Prop_2$.
We denote by $\Sig$ the category of signatures in $\HDQL$.

\subsection{Models}
A \emph{quantum model} over a signature $\Delta$ is a Kripke structure $(W,M)$ such that:
\begin{enumerate}
\item $W$ is  a first-order model of the theory $(\Sigma,E)$ such that 
\begin{enumerate}
\item ${W\red_{\Sigma^\hil}}$ is a Hilbert space, where ${W\red_{\Sigma^\hil}}$ is the reduct of $W$ to the first-order signature of Hilbert spaces $\Sigma^\hil$,
\item for all symbols $\ut:\vc\to\vc\in U$, the function $\ut^W:W_\vc\to W_\vc$ is a unitary transformation, that is, 
$\ut^W$ is a bounded linear operation which has an adjoint $(\ut^W)^\dagger$ that is its inverse, $\ut^W ; (\ut^W)^\dagger = (\ut^W)^\dagger ; \ut^W = 1_\W$,
\item for all symbols $q:\vc\to\vc\in Q$, the function $q^W:W_\vc\to W_\vc$ is a projective measurement, that is, there exists a closed subspace $\X\subseteq W_\vc$ such that $q^W(w)=P_\X(w)/\sqrt{\ip{w}{P_\X(w)}}$ for all vectors $w\in W_\vc$, 
where the function $P_\X:W_\vc\to W_\vc$ is the projection on $\X$.
\end{enumerate}

\item $M:W_\vc\to |\Mod^\PL(\Prop)|$ is a mapping from the set of vectors $W_\vc$ to the class of propositional models $|\Mod^\PL(\Prop)|$ s.t.
$r^{(W,M)}=\{w\in W_\vc \mid r\in M_w\}$
is a closed subspace for all closed propositional symbols $r\in \Prop_c$.~\footnote{Notice that the propositional logic model $M_w$ at the state $w$ consists of a set of propositional symbols from $\Prop$.}
\end{enumerate}
We let $(W,M)$ and $(W^i,M^i)$ range over quantum models, where $i$ is any index.

A homomorphism $h:(W,M)\to(W',M')$ in $\HDQL$ is a first-order homomorphism $h:W\to W'$ such that:
\begin{enumerate}
\item $h_\cn:W_\cn\to W'_\cn$ is the identity on complex numbers, and 
\item $M_w\subseteq M'_{h(w)}$ for all vectors $w\in W_\vc$.
\end{enumerate}
The proof of the following lemma is known.
\begin{lemma} \label{lemma:inj}
All homomorphisms of Hilbert spaces are injective. 
In particular, all homomorphisms of quantum models are injective.
\end{lemma}
\begin{remark}
The class of quantum models over a signature $\Delta$ together with their homomorphisms forms a category denoted $\Mod(\Delta)$.
\end{remark}
Given a signature morphism $\chi:\Delta\to\Delta'$,
the reduct ${(W',M')\red_\chi}$ of a  $\Delta'$-model $(W',M')$ is a $\Delta$-model defined by ${(W',M')\red_\chi}=(W,M)$, where
\begin{enumerate}
\item $W$ is the reduct of $W'$ across $\chi:\Sigma\to \Sigma'$ in first-order logic, in symbols, $W={W'\red_\chi}$, and
\item $M_w=\{p\in\Prop \mid \chi(p)\in M'_w\}$ is the reduct of $M'_w$ across $\chi:\Prop\to\Prop'$ in propositional logic, in symbols,
$M_w=M'_w\red_\chi$, for all vectors $w\in W'_\vc$.
\end{enumerate}
The reduct ${h'\red_\chi}$ of a homomorphism $h'\in\Mod(\Delta')$ is defined by $(h'\red_\chi)_\vc=h'_\vc$ and  $(h'\red_\chi)_\cn=h'_\cn$.
\begin{remark}
For each signature morphism $\chi:\Delta\to\Delta'$ in $\Sig$,
the model reduct ${\_\red_\chi}:\Mod(\Delta')\to\Mod(\Delta)$ is a functor.
Moreover,  $\Mod:\Sig\to\Cat^{op}$ defined by $\Mod(\chi)(h')=h'\red_\chi$ for all signature morphisms $\chi:\Delta\to\Delta'$ and all homomorphisms $h'\in\Mod(\Delta')$, is a functor.  
\end{remark}
\subsection{Sentences} \label{sec:sen}
The set of \emph{actions} over a signature $\Delta$ is defined by the following grammar:
\begin{center}
$\act\Coloneqq u \mid q \mid \act \comp \act \mid \act \cup \act \mid \act^{*}$,
\end{center}
where $u$ is a unitary transformation symbol and $q$ is a quantum measurement symbol.
We let $\act$ and $\act_i$ range over actions, where $i$ is any index.

The set of \emph{sentences}, $\Sen(\Delta)$, is defined by the following grammar:
\begin{center}
$\gamma \Coloneqq
p \mid
\at{k}\gamma \mid
\gamma \wedge \gamma \mid
\neg\gamma\mid
\sim \gamma \mid
\nec{\act}\gamma \mid
\store{z}\gamma$,
\end{center}
where $p$ is a propositional symbol,
$k$ is a term of sort vector,
$\act$ is an action, and 
$z$ is a variable of sort vector.
We refer to the sentence operators, in order, as 
\emph{retrieve},
\emph{conjunction},
\emph{negation},
\emph{quantum negation},
\emph{necessity}, and
\emph{store},
respectively.
Other quantum operators can be introduced as abbreviations.
For example, quantum disjunction $\gamma_1\oplus\gamma_2$ is defined by $\sim(\sim\gamma_1\wedge \sim\gamma_2)$,
for all sentences $\gamma_1$ and $\gamma_2$.
We make the following notational conventions:
\begin{itemize}
\item We let $k$ and $k_i$ range over terms of sort vector, where $i$ is any index.
\item We let $\gamma$ and $\gamma_i$ range over hybrid-dynamic sentences, where $i$ is any index.
\item Similarly, we let $\Gamma$ and $\Gamma'$ range over sets of hybrid-dynamic sentences.
\end{itemize}
Each signature morphism $\chi:\Delta_1\to \Delta_2$ induces a sentence translation $\chi:\Sen(\Delta_1) \to\Sen(\Delta_2)$ that replaces, in an inductive manner, in any sentence $\gamma\in\Sen(\Delta_1)$ the symbols from $\Delta_1$ with symbols from $\Delta_2$ according to $\chi:\Delta_1\to\Delta_2$. 
As in ordinary multi-modal hybrid propositional logic,
for each action $\act$,
one can define an operator \emph{until}~\cite{DBLP:journals/jsyml/ArecesBM01}:
\begin{center}
$Until_\act(\gamma_1,\gamma_2)\coloneqq\store{x}\pos{\act} \store{y} \gamma_1 \wedge \at{x} (\nec{\act}(\pos{\act} y \Rightarrow \gamma_2))$.
\end{center}
The current state is named $x$ and then 
$\pos{\act}$ is used to move to an accessible state, which is named $y$.
The first argument of the conjunction says that the sentence $\gamma_1$ holds in the state $y$.
The second argument of the conjunction sets the current state to $x$ by applying $\at{x}$;
then the sentence $\gamma_2$ holds in any state which succeeds  $x$ and precedes $y$.

\begin{remark}
The mapping $\Sen:\Sig\to \Set$ from the category of signatures $\Sig$ to the category of sets $\Set$ is a functor.  
\end{remark}
\subsection{Local satisfaction relation}
Let $(W,M)$ be a quantum model over a signature $\Delta$.
The semantics of actions is defined in the standard way:
\begin{enumerate*}[label=(\alph*)]
\item $(\act_1\comp\act_2)^W=\act_1^W\comp \act_2^W$;
\item $(\act_1\cup\act_2)^W=\act_1^W\cup \act_2^W$;
\item $(\act^*)^W=\bigcup_{n\in \N} (\act^n)^W$,
where $\act^0$ denotes the identity, and $\act^{n+1}=\act\comp \act^n$ for all natural numbers $n\in \N$.
\end{enumerate*}
The semantics of sentences is defined as follows:
\begin{itemize}
\item $p^{(W,M)}=\{w\in W_\vc\mid p\in M_w\}$ for all propositional symbols $p$ in $\Delta$;
\item $(\at{k}\gamma)^{(W,M)}=
\left\{\begin{array}{l l}
W_\vc & \text{if } k^W\in \gamma^{(W,M)}, \\
\emptyset  & \text{if } k^W\not\in \gamma^{(W,M)};
\end{array}\right.$
\item $(\gamma_1\wedge\gamma_2)^{(W,M)}= \gamma_1^{(W,M)} \cap \gamma_2^{(W,M)}$;
\item $(\neg\gamma)^{(W,M)}=W_\vc \setminus \gamma^{(W,M)}$;
\item $(\sim\gamma)^{(W,M)}= (\gamma^{(W,M)})^\perp$;
\item $(\nec{\act}\gamma)^{(W,M)}= \{w\in W_\vc\mid \act^W(w)\subseteq \gamma^{(W,M)}\}$, 

where $\act^W(w)=\{v\in W_\vc\mid (w,v)\in \act^W\}$;
\item $(\store{z}\gamma)^{(W,M)}=\{w\in W_\vc \mid w\in \gamma^{(W^{z\leftarrow w},M)} \}$,
where $(W^{z\leftarrow w},M)$ is the unique expansion of $(W,M)$ to $\Delta[z]$ which interprets $z$ as $w$.
Notice that $(W^{z\leftarrow w},M)$ interprets all symbols in $\Delta$ as $(W,M)$.
\end{itemize}
We say that $(W,M)$ \emph{satisfies $\gamma$ in the state $w$}, in symbols, $(W,M)\models^w \gamma$, if $w\in \gamma^{(W,M)}$. 
The following result shows that the truth is invariant w.r.t. the change of notation, that is, $\HDQL$ is a stratified institution~\cite{aig-strat}.
\begin{theorem}[Local satisfaction condition] \label{prop:sat-cond}
For all signature morphisms $\chi:\Delta\to\Delta'$, 
all quantum models $(W',M')$ defined over the signature $\Delta'$, 
all sentences $\gamma$ defined over the signature $\Delta$, and 
all vectors $w\in W'_\vc$, we have: 
$(W',M')\models^w \chi(\gamma)$ iff $(W',M')\red_\chi\models^w\gamma$.
\end{theorem}
Let $(W,M)$ be the reduct of $(W',M')$ across $\chi:\Delta\to\Delta'$.
Since $\chi$ is the identity on the signature of Hilbert spaces $\Sigma^\hil$, 
the Hilbert spaces ${W\red_{\Sigma^\hil}}$ and ${W'\red_{\Sigma^\hil}}$ coincide.
It follows that $W_\vc=W'_\vc$ which means that the local satisfaction condition is well-defined.

\subsection{Global satisfaction relation}
The global satisfaction relation between models and sentences is defined below:
\begin{itemize}
\item $(W,M)\models \gamma$, read $(W,M)$ globally satisfies $\gamma$, when $\gamma^{(W,M)}=W_\vc$.
\end{itemize}
In formal methods, the global satisfaction relation is at the core of formal verification, 
since the engineers need to model software and hardware systems with sets of sentences that need to be satisfied globally.
The global satisfaction relation between models and sentences is naturally extended to a satisfaction relation between sets of sentences.
\begin{itemize}
\item $\Gamma\models\gamma$, read $\Gamma$ globally satisfies $\gamma$, when $(W,M)\models\Gamma$ implies $(W,M)\models\gamma$, 
for all quantum models $(W,M)$.
\end{itemize}
Notice that $\emptyset\models\bigwedge\Gamma\Rightarrow\gamma$ implies $\Gamma\models\gamma$  but the backward implication does not hold.
This means that the semantics of the satisfaction relation $\models$ between sentences is different from the standard one used in modal logic literature.
\begin{itemize}
\item $\Gamma\models^k\gamma$, read $\Gamma$ satisfies $\gamma$ at $k$, when $(W,M)\models\Gamma$ implies $(W,M)\models^{(k^W)} \gamma$, for all models $(W,M)$.
\end{itemize}
Notice that $\Gamma\models^k\gamma$ iff $\Gamma\models\at{k}\gamma$. 
Also, $\emptyset\models^k\bigwedge\Gamma\Rightarrow\gamma$ implies $\Gamma\models^k\gamma$  but the backward implication does not hold.
The following result is a direct corollary of Theorem~\ref{prop:sat-cond}.
\begin{corollary} \label{cor:trans}
$\Gamma\models^k\gamma$ implies $\chi(\Gamma)\models^{\chi(k)}\chi(\gamma)$,
for all signature morphisms $\chi:\Delta\to\Delta'$, all sets of $\Delta$-sentences $\Gamma$ and all $\Delta$-sentences~$\gamma$.
\end{corollary}

\subsection{Closed sentences}
In quantum logic literature, a distinguished class of sentences consist of all sentences that are interpreted as closed spaces. In this section, we will briefly look into some of their properties. The results presented in this subsection are from~\cite{gai-qinit}.
The set of \emph{closed sentences} $\Sen_c(\Delta)$ over a signature $\Delta$ is defined by the following grammar:
\begin{center}
$\rho\Coloneqq 
r\mid
\sim\rho\mid
\rho\wedge\rho\mid 
\nec{\bact}\rho$,
\end{center}
where $r$ is a closed propositional symbol, $\bact$ is a unitary action, that is, an action free of quantum measurement symbols.
The following result shows that the interpretation of any closed sentence in a model is a closed subspace.
\begin{theorem}\label{th:Hibert-sat}
The semantics of any closed sentence is a closed subspace, i.e.,
for all models $(W,M)$ and all closed sentences $\rho$ defined over the same signature, 
$\rho^{(W,M)}$ is a closed subspace.
\end{theorem}
Closed sentences have some unique features which distinguish them from the rest of the sentences.
Some properties of closed sentences are stated in the next corollary.
\begin{corollary} \label{cor:Hilbert-sat}
For all models $(W,M)$ and all closed sentences $\rho$ defined over the same signature, we have:
\begin{enumerate}
\item $(W,M)\models^{(0^W)} \rho$.

\item If $(W,M)\models^w \rho$ then $(W,M)\models^{a w} \rho$, 
for all vectors $w\in W_\vc$ and all complex numbers $a\in W_\cn$.

\item If $(W,M)\models^{w_1} \rho$ and $(W,M)\models^{w_2} \rho$ then $(W,M)\models^{w_1 + w_2} \rho$, 
for all vectors $w_1,w_2\in W_\vc$.

\item Let $\{w_n\}_{n\in \N}$ be a Cauchy sequence of vectors, and let $w$ be its limit.

If $(W,M)\models^{w_n} \rho$ for all $n\in \N$ then $(W,M)\models^{w} \rho$.
\end{enumerate}
\end{corollary}
Sasaki hook $\rho_1\leadsto \rho_2$ is defined by $\sim(\rho_1 \wedge \sim(\rho_1\wedge\rho_2))$,
for all closed sentences $\rho_1$ and $\rho_2$.
The following lemma shows that Sasaki hook can be viewed as an implication for closed sentences, that is, a quantum implication.
\begin{lemma}[Quantum implication] \label{lemma:hook} 
For all quantum models $(W,M)$ and all closed sentences $\rho_1,\rho_2$ defined over the same signature, we have:
\begin{enumerate}
\item $\rho_1^{(W,M)}\cap(\rho_1\leadsto\rho_2)^{(W,M)}\subseteq \rho_2^{(W,M)}$ and
\item $\rho_1^{(W,M)}\subseteq \rho_2^{(W,M)}$ iff $(W,M)\models \rho_1\leadsto \rho_2$.
\end{enumerate}
\end{lemma}
\subsection{Entailment systems}
The syntactic counterpart of the satisfaction relation for reasoning about the logical consequences of sentences is defined below.

\begin{figure}[h]\centering
\begin{tabular}{l l}
&  \\
$(Monotonicity)~\Frac{\gamma\in\Gamma}{\Gamma\vdash^k \gamma}$  & 
$(Unions)~\Frac{\Gamma\vdash^k\gamma}{\Gamma\cup\Gamma'\vdash^k\gamma}$\\ 
&\\
$(Cut)~\Frac{\Gamma\vdash^k\gamma_2 \Space \Gamma\cup \{ \gamma_2\} \vdash^k \gamma_1}{\Gamma\vdash^k\gamma_1}$ &
$(Translation)~\Frac{\Gamma \vdash^k \gamma}{\chi(\Gamma)\vdash^{\chi(k)}\chi(\gamma)}$ where $\chi:\Delta\to\Delta'$ \\ 
& \\
\end{tabular}
\caption{Entailment systems}
\label{table:entail}
\end{figure}

\begin{definition}[Entailment systems]
An \emph{entailment system}  is a family of relations $\vdash=\{\vdash^k_\Delta\}_{\Delta\in|\Sig|,k\in T_\Sigma}$ between sets of sentences and sentences (that is, $\vdash_\Delta^k\subseteq \mathcal{P}(\Sen(\Delta))\times \Sen(\Delta)$ for all signatures $\Delta$ and all terms $k$ of sort vector) satisfying $(Monotonicity)$, $(Unions)$ and $(Translation)$ defined in Fig.~\ref{table:entail}.
\end{definition}

We drop the subscript $\Delta$ from the notion $\vdash_\Delta^k$ when there is no danger of confusion.
$(Cut)$ is omitted from the definition of entailment system which shows that lemma discovery is not needed for the proof calculus developed in this paper.

\begin{definition}[Entailment properties]
Let $\vdash$ be an entailment system.
\begin{enumerate}
\item $\vdash$ is \emph{sound (complete)} if for all signatures $\Delta$ and all terms $k$, we have: 
$\vdash_\Delta^k \ \subseteq\ \models_\Delta^k$ ($\models_\Delta^k \ \subseteq\ \vdash_\Delta^k$).

\item $\vdash$ is \emph{compact} if for all signatures $\Delta$, all sets of sentences $\Gamma$, all sentences $\gamma$, and all terms $k$, we have:
$\Gamma\vdash_\Delta^k\gamma$ implies $\Gamma_f\vdash_\Delta^k\gamma$ for some finite $\Gamma_f\subseteq\Gamma$.
\end{enumerate}
\end{definition}
\section{Birkhoff completeness} \label{sec:birkhoff}
We introduce two classes of sentences for which we study soundness, compactness and completeness.
\begin{definition}[Clauses]
Let $\Delta$ be any signature in $\HDQL$.
\begin{enumerate}
\item The set of \emph{basic sentences} over $\Delta$ is defined by the following grammar:
\begin{center}
$\varphi \Coloneqq
p \mid
\varphi \wedge \varphi \mid
\at{k}\varphi \mid
\nec{\act}\varphi\mid
\store{z}\varphi$,
\end{center}
where $p$ is a propositional symbol,
$k$ is a term of sort vector, 
$\act$ is an action, and
$z$ is a variable of sort vector.
\item The set of \emph{quantum clauses} over $\Delta$ is defined by the following grammar:
\begin{center}
$\gamma \Coloneqq
p \mid
\rho_1\leadsto\rho_2\mid
\varphi\Rightarrow \gamma\mid
\gamma\wedge\gamma\mid
\at{k}\gamma\mid
\nec{\act}\gamma \mid 
\store{z}\gamma$,
\end{center}
where $p$ is a propositional symbol,
$\rho_1$ is a closed basic sentence, 
$\rho_2$ is a closed quantum clause, 
$\varphi$ is a basic sentence,
$k$ is a term of sort vector, 
$\act$ is an action, and
$z$ is a variable of sort vector.
\end{enumerate}
\end{definition}

\subsection{Logical framework $\mathcal{L}$} \label{sec:L}
We define the logical framework in which the results will be proved.
This is a fragment of $\HDQL$ defined in Section \ref{sec:HDQL}.

Let $\Delta=(\Sigma,\emptyset,\Prop)$ be a signature such that $\Sigma=(S^\hil,F^\hil\cup U\cup Q,P^\hil)$, 
where $\Sigma^\hil=(S^\hil,F^\hil,P^\hil)$ is the signature of Hilbert spaces defined in Section~\ref{sec:HDQL}, 
$U$ is a set of unitary transformation symbols and 
$Q$ is a set of measurement symbols.
Let $\W$ be a first-order model defined over the signature $\Sigma$ such that 
\begin{enumerate}
\item $\W\red_{\Sigma^\hil}$ is a Hilbert space,
\item $u^\W:\W_\vc\to\W_\vc$ is a unitary transformation for all $u:\vc\to\vc\in U$, and
\item $q^\W:\W_\vc\to\W_\vc$ is a quantum measurement for all $q:\vc\to\vc\in Q$.
\end{enumerate}
We make the following notational conventions:
\begin{enumerate} 
\item Let $\Sigma_\W$ be the first-order signature obtained from $\Sigma$ by adding all elements in $\W$ as constants, that is, 
$\Sigma_\W=(S^\hil,F^\hil\cup U \cup Q\cup D_\W \cup C_\W, P^\hil)$, where
\begin{enumerate}
\item $D_\W=\W_\vc$, the set of vectors in $\W$, and 
\item $C_\W=\mathbb{C}$, the set of complex numbers.
\end{enumerate}
\item Let $\W_\W$ be the first-order model over $\Sigma_\W$ obtained from $\W$ by interpreting each constant $c\in C_\W$ as the complex number $c$ and each constant $w\in D_\W$ as the vector $w$.
\item Let $E_\W$ be the set of (ground) equations and relations satisfied by $\W_\W$, which means that $(\Sigma_\W,E_\W)$ is the positive diagram of $\W$.
\end{enumerate}
In classical model theory, it is well-known that $\Sigma_\W$-models which satisfy $E_\W$ are in one-to-one correspondence with the $\Sigma$-homomorphisms with the domain $\W$, that is, 
there is an isomorphism of categories $\Mod(\Sigma_\W,E_\W)\cong \W/\Mod(\Sigma)$.

The underlying logic in which the subsequent results will be developed is an arbitrary fragment $\mathcal{L}$ of $\HDQL$ satisfying the following two properties:
\begin{enumerate} 
\item All signatures are of the form $(\Sigma_\W,E_\W,\Prop)$ as described above. 
\item Retrieve $@$ belongs to the vocabulary of $\mathcal{L}$ if classical implication $\Rightarrow$ or quantum implication $\leadsto$ belongs to the vocabulary of $\mathcal{L}$.
\end{enumerate}
The fragment $\mathcal{L}$ is obtained from $\HDQL$ by restricting its syntax, that is, its signatures and sentences. 
Given a signature morphism $\chi:(\Sigma_\W,E_\W,\Prop) \to (\Sigma'_{\W'},E_{\W'},\Prop')$ in $\mathcal{L}$, 
since $(\Sigma_\W,E_\W)$ and $(\Sigma'_{\W'},E_{\W'})$ are the positive diagrams of the first-order models $\W$ and $\W'$, respectively,
by Lemma~\ref{lemma:inj},
the first-order signature morphism  $\chi:\Sigma_\W \to \Sigma'_{\W'}$ is injective.
Moreover, $\chi:\Sigma_\W \to \Sigma'_{\W'}$ is the identity on complex numbers.
In addition, one or more sentence operators from $\HDQL$ can be discarded. 
Of course, if retrieve is discarded from the grammar used to define sentences in $\HDQL$ then both classical and quantum implication must be dropped.
This condition is necessary because the proof rules for both classical and quantum implication depend on the existence of retrieve.
See Section~\ref{sec:horn}.
Other than that, the sentence operators are independent.
For example, quantum implication is not required by any of the results in this paper.
If quantum implication occurs, there are proof rules and arguments to deal with it. 
If quantum implication is not part of the vocabulary then the proofs still hold, since the cases corresponding to quantum implication can be, simply, discarded.
$\mathcal{L}$ can be regarded as a parameter for the subsequent developments which can be adjusted depending on the applications.
\begin{figure}[h]\centering
\begin{tabular}{l l}
& \\
$(Origin)~\Frac{}{\Gamma\vdash^{0} r}$ &
$(Mult)~\Frac{\Gamma\vdash^w r}{\Gamma\vdash^{a w} r}$ \\
& \\
$(Add)~\Frac{\Gamma\vdash^{w_1}r \Space \Gamma\vdash^{w_2}r}{\Gamma\vdash^{w_1 + w_2}r}$ \hspace{0.7cm} &
$(Cauchy)~\Frac{ \Gamma\vdash^{w_n} r \text{ for all } n\in \mathbb{N}}{\Gamma\vdash^{w}r }$ \\
&\\
where $\{w_n\}_{n\in \mathbb{N}}$ \rlap{is a Cauchy sequence such that  $\lim_{n\to \infty} w_n=w$}\\
&\\
\end{tabular}
\caption{Closed propositional symbols}
\label{table:closed}
\end{figure}
Fig.~\ref{table:closed} contains the proof rules for reasoning about closed propositions.
Notice that $r$ ranges over closed propositional symbols, 
$w$ ranges over vectors and $a$ ranges over complex numbers.
$(Cauchy)$ is an infinitary rule, since it has a countably infinite number of premises, one for each natural number $n\in\mathbb{N}$.
In applications, it is rarely used.
Assume, for example, that each closed propositional symbol $r$ is defined by an orthonormal basis $\{v_1,\dots,v_n\}$ of a closed subspace, that is, 
$\{\at{v_1}r,\dots,\at{v_n}r\}\subseteq \Gamma$ and $r$ does not occur positively in any of the sentences from $\Gamma\setminus \{\at{v_1}r,\dots,\at{v_n}r\}$. However, $r$ can occur, for example, in the conditional part of the clauses from $\Gamma\setminus \{\at{v_1}r,\dots,\at{v_n}r\}$.
In this case, $(Cauchy)$ can be discarded because the set of states where each $r$ holds is the closed subspace generated by linear combinations of vectors from $\{v_1,\dots,v_n\}$.
\begin{definition} [Models defined by sentences] \label{def:init} 
Let $\vdash$ be any sound entailment system of $\mathcal{L}$ closed under the proof rules defined in Fig.~\ref{table:closed}.
Any set of sentences $\Gamma$ over a signature $\Delta_\W$ defines a model $(W^\Gamma,M^\Gamma)$ as follows:
\begin{enumerate}
\item $W^\Gamma = \W_\W$, and 
\item $M^\Gamma:W^\Gamma_\vc\to |\Mod^\PL(\Sigma_\W)|$ is defined by
$M^\Gamma_w=\{p\in\Prop\mid \Gamma\vdash^w p\}$ for all vectors $w\in W^\Gamma_\vc$.
\end{enumerate}
\end{definition}
The proof rules from Fig.~\ref{table:closed} ensure that $(W^\Gamma,M^\Gamma)$ is well-defined.
We will show that $(W^\Gamma,M^\Gamma)$ is the initial model of $\Gamma$ if $\Gamma$ is a set of quantum clauses.

\subsection{Basic sentences}\label{sec:basic}
In this subsection, we define proof rules for reasoning about basic sentences and then we prove their completeness.
\begin{figure}[h]
\begin{tabular}{l l}
& \\
$(EQ)~\Frac{\Gamma\vdash^{k_1} \gamma}{\Gamma\vdash^{k_2} \gamma}$ & for all $k_1=k_2\in E_\W$ \\
& \\
$(Ret_I)~\Frac{\Gamma\vdash^{k_1}\gamma}{\Gamma\vdash^{k_2}\at{k_1}\gamma}$ & 
$(Ret_E)~\Frac{\Gamma\vdash^{k_2}\at{k_1}\gamma}{\Gamma\vdash^{k_1}\gamma}$ \\
& \\
$(Store_I)~\Frac{\Gamma\vdash^k\gamma[z\leftarrow k]}{\Gamma\vdash^k\store{z}\gamma}$ & 
$(Store_E)~\Frac{\Gamma\vdash^k\store{z}\gamma}{\Gamma\vdash^k\gamma[z\leftarrow k]}$\\
& \\
$(Conj_I)~\Frac{\Gamma\vdash^k\gamma_1 \Space \Gamma\vdash^k\gamma_2}{\Gamma\vdash^k\gamma_1 \wedge \gamma_2}$ \hspace{1cm} & 
$(Conj_E)~\Frac{\Gamma\vdash^k\gamma_1 \wedge \gamma_2}{\Gamma\vdash^k\gamma_i}$ for all $i\in\{1,2\}$\\
& \\   
\end{tabular}
\caption{Basic proof rules}
\label{table:basic}
\end{figure}
$(EQ)$ defined in Fig.~\ref{table:basic} says that for each equation $k_1=k_2$ in $E_\W$, if $\Gamma\vdash^{k_1}\gamma$ holds then one can deduce $\Gamma\vdash^{k_2}\gamma$.
In practice, it is necessary to have an efficient way to compute projective measurements and unitary transformations for establishing the validity of $k_1=k_2$.
See Section~\ref{sec:case-study} for a concrete example.
The sentence $\gamma[z\leftarrow k]$ used to define $(Store_I)$ and $(Store_E)$ in Fig.~\ref{table:basic} is obtained from $\gamma$ by substituting the term $k$ for the variable $z$.
\begin{figure}[h]
\begin{tabular}{l l}
& \\
$(FT_I)~\Frac{\Gamma\vdash^{f(k)}\gamma}{\Gamma\vdash^k \nec{f}\gamma}$ &  
$(FT_E)~\Frac{\Gamma\vdash^k \nec{f}\gamma}{\Gamma\vdash^{f(k)}\gamma}$ 
 for all $f\in U\cup Q$ \\
&\\
$(Comp_I)~\Frac{\Gamma\vdash^k \nec{\act_1\comp\act_2}\gamma}{\Gamma\vdash^k \nec{\act_1}\nec{\act_2}\gamma}$ & 
$(Comp_E)~\Frac{\Gamma\vdash^k \nec{\act_1}\nec{\act_2}\gamma}{\Gamma\vdash^k \nec{\act_1\comp\act_2}\gamma}$ \\
& \\
$(Union_I)~\Frac{\Gamma\vdash^k \nec{\act_1}\gamma \Space \Gamma\vdash^k \nec{\act_2}\gamma}{\Gamma\vdash^k \nec{\act_1\cup\act_2}\gamma}$ & 
$(Union_E)~\Frac{\Gamma\vdash^k \nec{\act_1\cup\act_2}\gamma}{\Gamma\vdash^k \nec{\act_i}\gamma}$ for all $i\in\{1,2\}$ \\
& \\ 
$(Star_I)~\Frac{\Gamma\vdash^k \nec{\act^n}\gamma \text{ for any } n\in \mathbb{N}}{\Gamma\vdash^k \nec{\act^*}\gamma}$ & 
$(Star_E)~\Frac{\Gamma\vdash^k \nec{\act^*}\gamma}{\Gamma\vdash^k \nec{\act^n}\gamma}$ for all $n\in\mathbb{N}$ \\
& \\ 
\end{tabular}
\caption{Necessity}
\label{table:nec}
\end{figure}
Notice that $(Star_I)$ is an infinitary proof rule, since it has a countably infinite number of premises, one for each natural number $n\in\mathbb{N}$.
One needs inductive arguments to ensure that the premises of $(Star_I)$ are satisfied.
However, if $*$ is missing from the grammar which defines actions in $\mathcal{L}$ then completeness still holds.

\begin{lemma}[Basic soundness]\label{lemma:basic-sound}
The least entailment system of $\mathcal{L}$ closed under the proof rules defined in Fig.~\ref{table:closed}~--~\ref{table:nec} is sound.
\end{lemma}
The proof of soundness relies on the closure of satisfaction relation under the proof rules defined in Fig.~\ref{table:closed}~--~\ref{table:nec}.
\begin{proposition}[Basic compactness] \label{prop:basic-comp}
Assume that the signatures of $\mathcal{L}$ have no closed propositional symbols, and  
the actions of  $\mathcal{L}$ are free of the operator $*$.
Then the least entailment system of $\mathcal{L}$ closed under the proof rules defined in Fig.~\ref{table:basic}~--~\ref{table:nec} is \emph{compact}.
\end{proposition}
Proposition~\ref{prop:basic-comp} establishes an important result, since it provides conditions for making deductions from a finite number of premises. 
The number of vectors is uncountable, but according to Proposition~\ref{prop:basic-comp}, for the formal proofs,
it is enough to know the result of applying unitary transformations and projective measurements to the vectors occurring in the underlying specification.

\begin{theorem}[Basic completeness] \label{th:basic}
Let $\vdash$ be any sound entailment system of $\mathcal{L}$ closed under the proof rules defined in Fig.~\ref{table:closed}~--~\ref{table:nec}.
Let $\varphi$ be a basic sentence defined over a signature $\Delta_\W$. 
Let $k$ be a term of sort vector defined over $\Sigma_\W$.
We denote by $w$ the interpretation of $k$ in $W^\Gamma$.
\begin{enumerate}
\item $\Gamma\vdash^k\varphi$ ~iff~ $( W^\Gamma,M^\Gamma)\models^w\varphi$,  
for all sets of sentences $\Gamma$.
\item $\Phi\vdash^k\varphi$ iff $\Phi\models^k\varphi$ ~iff~  $(W^\Phi,M^\Phi)\models^w\varphi$,
for all sets of basic sentences $\Phi$.
\end{enumerate}
\end{theorem}
The sentences in $\Gamma$ from the first statement of Theorem~\ref{th:basic} are not necessarily basic.
An application of the first statement is obtained in the following subsection where $\Gamma$ is instantiated by a set of quantum clauses.
The second statement of Theorem~\ref{th:basic} is a direct consequence of the first due to the fact that $(W^\Phi,M^\Phi)$ satisfies globally $\Phi$.
It says that in order to reason about basic sentences one needs to consider only its initial models.

\subsection{Quantum clauses} \label{sec:horn}
In this section, we define proof rules for reasoning about quantum clauses and then we prove their completeness.

\begin{figure}[h]\centering
\begin{tabular}{l l} 
& \\
$(MP)~\Frac{\Gamma\vdash^k \varphi\Rightarrow \gamma \Space \Gamma\vdash^k \varphi}{\Gamma\vdash^k \gamma}$ \hspace{1cm} & 
$(MP_c)~\Frac{\Gamma\vdash^k \rho_1\leadsto \rho_2 \Space \Gamma\vdash^k \rho_1}{\Gamma\vdash^k \rho_2}$ \\
& \\
$(Imp)~\Frac{\Gamma\cup\{\at{k}\varphi\}\vdash^k\gamma}{\Gamma\vdash^k\varphi\Rightarrow \gamma}$ &
$(Imp_c)~\Frac{\Gamma\cup\{\at{k}\rho_1\}\vdash^k\rho_2}{\Gamma\vdash^k\rho_1\leadsto \rho_2}$ \\
& \\
\end{tabular}
\caption{Implication}
\label{table:imp}
\end{figure}

The proof rules for implication are defined in Fig.~\ref{table:imp}.
Recall that retrieve $@$ belongs to the vocabulary of $\mathcal{L}$ if classical implication $\Rightarrow$ or quantum implication $\leadsto$ belongs to the vocabulary of $\mathcal{L}$.
Therefore, $(Imp)$ and $(Imp_c)$ are well-defined.

\begin{lemma}[Soundness]\label{lemma:birkhoff-sound}
The least entailment system of $\mathcal{L}$ closed under the proof rules defined in Fig.~\ref{table:closed}~--~\ref{table:imp} is sound.
\end{lemma}
As usual, the proof of soundness is based on the closure of satisfaction relation under the proof rules defined in Fig.~\ref{table:closed}~--~\ref{table:imp}.
\begin{proposition}[Compactness]
Assume that the signatures of $\mathcal{L}$ have no closed propositional symbols, and  
the actions of  $\mathcal{L}$ are free of the operator $*$.
Then the least entailment system of $\mathcal{L}$ closed under the proof rules defined in Fig.~\ref{table:basic}~--~\ref{table:imp} is \emph{compact}.
\end{proposition}
Compactness for quantum clauses holds under the same conditions as for basic sentences.
Initiality is proved in \cite{gai-qinit} using semantic arguments, while in the present contribution the proof of initiality relies on syntactic arguments based on proof rules.
\begin{theorem} [Initiality]\label{th:init}
Let $\vdash$ be any sound entailment system of $\mathcal{L}$ closed under the proof rules from Fig.~\ref{table:closed}~--~\ref{table:imp}.
For all sets of quantum clauses $\Gamma$, all quantum clauses $\gamma$, all basic sentences $\varphi$ and all terms $k$, we have:
\begin{enumerate}
\item $\Gamma\vdash^k\gamma$ implies $( W^\Gamma,M^\Gamma)\models^w \gamma$, 
where $w=k^{(W^\Gamma)}$. \newline
In particular, $(W^\Gamma,M^\Gamma)$ is the initial model of $\Gamma$.

\item $\Gamma\vdash^k\varphi$ ~iff~ $\Gamma\models^k\varphi$ ~iff~ $(W^\Gamma,M^\Gamma)\models^w\varphi$,
where $w=k^{(W^\Gamma)}$.
\end{enumerate}
\end{theorem}
The first statement of Theorem~\ref{th:init} shows that any set of quantum clauses has an initial model which, in particular, means that any set of quantum clauses is satisfiable.
The second statement of Theorem~\ref{th:init} shows that the satisfaction of basic sentences by a set of quantum clauses can be established by checking if the satisfaction holds in the initial model.
It is worth mentioning that the proof of Theorem~\ref{th:init} does not rely on $(Imp)$ and $(Imp_c)$.
The proof rules $(Imp)$ and $(Imp_c)$ are needed for the proof of completeness. 
\begin{theorem}[Completeness] \label{th:birkhoff}
Let $\vdash$ be any sound entailment system of $\mathcal{L}$ closed under the proof rules defined in Fig.~\ref{table:closed}~--~\ref{table:imp}.
For all sets of quantum clauses $\Gamma$,
all quantum clauses $\gamma$ and 
all terms $k$, we have
$\Gamma\models^k\gamma$ ~iff~ $\Gamma\vdash^k\gamma$.
\end{theorem}

\section{A case study} \label{sec:case-study}
We show the practicality of the proof-theoretic infrastructure developed above.
Working within a positive diagram of a Hilbert space means in practice that there is a library defining scalars, vectors and the operations on them which is available to the engineers to specify quantum programs.
The projection on a closed space $\mathcal{X}$ with an orthonormal basis $\{v_1,\dots,v_n\}$, $P_\mathcal{X}:\W_\vc\to\mathcal{X}$, is defined by $P_\mathcal{X}(w)= a_1 v_1 + \dots + a_n v_n$, where $a_i=\ip{v_i}{w}$ for all $i\in\{1,\dots,n\}$.
Therefore, projective measurements are not difficult to define once an orthonormal basis for the corresponding closed space is provided.
There are many examples of unitary transformations which are called quantum gates.
In practice, all quantum gates used to define quantum circuits are obtained from the single qubit gates and the controlled-NOT gate in the 2-qubit system,
$\mathsf{CNOT}=\begin{pmatrix}
1 & 0 & 0 & 0\\
0 & 1 & 0 & 0\\
0 & 0 & 0 & 1\\
0 & 0 & 1 & 0
\end{pmatrix}$.
Examples of frequently used single qubit gates are 
the Hadamard gate 
$\mathsf{H}=\displaystyle
\frac{1}{\sqrt{2}}
\begin{pmatrix}
1 & 1 \\
1 & -1
\end{pmatrix}$
and the Pauli gates
$\mathsf{X}=
\begin{pmatrix}
0 & 1 \\
1 & 0
\end{pmatrix}$, 
$\mathsf{Y}=
\begin{pmatrix}
0 & -i \\
i & 0
\end{pmatrix}$
 and 
$\mathsf{Z}=
\begin{pmatrix}
1 & 0 \\
0 & -1
\end{pmatrix}$.
Therefore, assuming that the quantum gates and the projective measurements are predefined, that is, working in a logical framework in which a positive diagram of a Hilbert space is given, is not a shortcoming.

\subsection{Quantum teleportation}
\begin{figure}[h]\centering
\begin{quantikz}
\lstick{\ket{\w}} & \ctrl{1} & \gate{H}  & \meter{} & \cw{i}  & \cwbend{2} \\
\lstick[wires=2]{$\ket{\beta_{00}}$} & \targ{} & \qw & \meter{} & \cwbend{1} \cw{j}   \\
  &  \qw  &   \qw   & \qw  & \gate{X^j} & \gate{Z^i} & \qw & \rstick{$\ket{\w}$}
\end{quantikz}
\caption{Quantum teleportation}
\label{fig:teleport}
\end{figure}
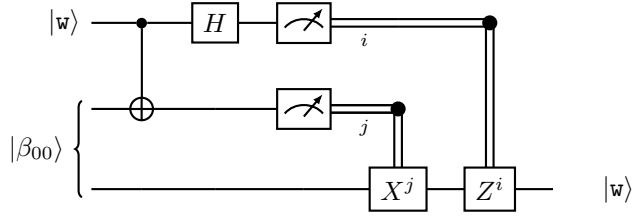

Quantum teleportation is a protocol for moving the state of a quantum system in the absence of a quantum communication channel linking a sender to a recipient.
The sender and the recipient, traditionally called Alice and Bob, are separated in space. 
Each has one qubit of $\ket{\beta_{00}}=\displaystyle\frac{\ket{00}+\ket{11}}{\sqrt{2}}$.
In addition to her part of $\beta_{00}$, Alice holds a qubit $\w=\alpha \ket{0} +\beta\ket{1}$, where $\alpha$ and $\beta$ are unknown amplitudes.
Alice ``teleports'' the qubit $\ket{\w}$ to Bob, that is, she performs a program that has the input $\ket{\w_0}=\ket{\w}\otimes \ket{\beta_{00}}$ and the output $\ket{i}\otimes\ket{j} \otimes \ket{\w}$, where $i,j\in\{0,1\}$.
The quantum circuit is depicted in Fig.~\ref{fig:teleport}.
The Hilbert space for this protocol is the $3$-qubit system $\Hil\otimes \Hil\otimes \Hil$, where $\Hil$ denotes the 2-dimensional Hilbert space.
The starting signature is $\Delta=(\Sigma,\emptyset,\Prop)$, where 
\begin{enumerate}
\item $\Sigma$ is obtained from the signature of Hilbert spaces $\Sigma^\hil$ by adding 
the set of unitary transformation symbols $U=\{ u_0,u_1,\sigma_0,\sigma_1,\delta_0,\delta_1\}$ and 
the set of measurement symbols $Q=\{q_{00},q_{01},q_{10},q_{11}\}$, and
\item $\Prop=\{p\}$.
\end{enumerate}
The first-order model $\W$ is obtained from the 3-qubit system $\Hil\otimes \Hil\otimes \Hil$ by interpreting 
\begin{enumerate}
\item the unitary transformation as follows:
\begin{enumerate*}
\item $u_0^\W= CNOT\otimes I_2$, 
\item $u_1^\W=H\otimes I_4$,
\item $\sigma_0^\W=I_4\otimes X^0$, 
\item $\sigma_1^\W=I_4 \otimes X^1$,
\item $\delta_0^\W=I_4\otimes Z^0$, 
\item $\delta_1^\W=I_4\otimes Z^1$; 
\end{enumerate*}

\item the quantum measurement symbols as follows:
$q_{ij}^\W$ is the measurement corresponding to the projection on the closed subspace generated by the vectors $\{\ket{ij}\otimes\ket{0},\ket{ij}\otimes\ket{1}\}$, where $i,j\in\{0,1\}$.
\end{enumerate}

Let $\Phi$ be the set of basic sentences $\{\at{\ket{ij}\otimes \ket{\w}} p \mid i,j\leq 1 \}$.
We formally verify that the output is $\ket{i}\otimes\ket{j}\otimes \ket{\w}$, that is, 
$(W^\Phi,M^\Phi)\models^{\ket{\w}\otimes \ket{\beta_{00}}}  \nec{ \bigcup_{i,j\in\{0,1\}} \act_{ij}} p$, 
where $\act_{ij}=u_0\comp u_1 \comp q_{ij} \comp \sigma_j \comp \delta_i$ for all $i,j\in\{0,1\}$.
By Theorem~\ref{th:init}, we have:
\begin{enumerate}
\item $(W^\Phi,M^\Phi)$ is the initial model of $\Phi$, and
\item $(W^\Phi,M^\Phi)\models^{\ket{\w}\otimes \ket{\beta_{00}}}  \nec{ \bigcup_{i,j\in\{0,1\}} \act_{ij}} p$ iff 
$\Phi\models^{\ket{\w}\otimes \ket{\beta_{00}}} \nec{\bigcup_{i,j\in\{0,1\}} \act_{ij}} p$.
\end{enumerate}
Notice that since $\at{\ket{ij}\otimes\ket{w}}p\in\Phi$, 
by $(Monotonicity)$, $\Phi\vdash^k \at{\ket{ij}\otimes\ket{w}}p$ for any term $k$ of sort vector. 
By $(Ret_E)$, $\Phi\vdash^{\ket{ij}\otimes\ket{w}}p$.
The following are equivalent:
\begin{proofsteps}{22em}
$\Phi\vdash^{\ket{\w}\otimes \ket{\beta_{00}}} \nec{\bigcup_{i,j\in\{0,1\}} \act_{ij}} p$ & 
by $(Union_E)$ and $(Union_I)$\\
$\Phi\vdash^{\ket{\w}\otimes \ket{\beta_{00}}} \nec{\act_{ij}} p$ for all $i,j\in\{0,1\}$ &
by the definition of $\act_{ij}$\\
$\Phi\vdash^{\ket{\w}\otimes \ket{\beta_{00}}}  \nec{u_0\comp u_1 \comp q_{ij} \comp \sigma_j \comp \delta_i} p$ for all $i,j\in\{0,1\}$ & 
by $(FT_E)$ and $(FT_I)$\\
$\Phi\vdash^{k_{ij}} p$,  
where $k_{ij}$ is the following term \newline $\delta_i(\sigma_j(q_{ij}(u_1(u_0(\ket{\w}\otimes \ket{\beta_{00}})))))$, 
for all $i,j\in\{0,1\}$ & 
by $(EQ)$, since $k_{ij}=\ket{ij}\otimes\ket{w} \in E_\W$ \\
$\Phi\vdash^{\ket{ij}\otimes\ket{w}}p$ for all $i,j\in\{0,1\}$ & which was proved above
\end{proofsteps}
\noindent Hence, $\Phi\vdash^{\ket{\w}\otimes \ket{\beta_{00}}} \nec{\bigcup_{i,j\in\{0,1\}} \act_{ij}} p$ holds.
We reiterate the comment made for $(EQ)$ in Section~\ref{sec:basic}: 
in order to establish the validity of $\delta_i(\sigma_j(q_{ij}(u_1(u_0(\ket{\w}\otimes \ket{\beta_{00}})))))=\ket{ij}\otimes\ket{w}$ one needs an efficient tool to compute the application of unitary transformations and projective measurements.

\section{Conclusions} \label{sec:conclusion}
The hybrid-dynamic quantum logic studied in this contribution is obtained by enriching hybrid-dynamic propositional logic using a two-layered approach to the design and analysis of quantum systems: 
\begin{enumerate*}[label=(\alph*)]
\item \emph{a local view} concerning the structural properties of the states which are vectors of a Hilbert space, and
\item \emph{a global view}, which corresponds to a specialized language for capturing the dynamics of quantum systems.
\end{enumerate*}
In this way, quantum systems can be modeled as Kripke structures whose frames are Hilbert spaces together with a set of unitary transformations and projective measurements.
We use Robinson diagrams to define the frames which allows formal methods practitioners to focus on the dynamics of quantum systems assuming that the vectors and the operations on them are already defined.
Similarly, quantum programs are built by relying on libraries that provide data types such as scalars and vectors.

We have developed a layered approach towards a Birkhoff completeness result:
first, the basic layer, which deals with entailments where both the premises and the conclusion are basic sentences;
second, a mixed layer, which deals with entailments where the premises are quantum clauses, but the conclusion is a basic sentence; and
third, a quantum clause layer, which deals with entailments where both the premises and the conclusion are quantum clauses.

The set of states is not the set of one-dimensional closed subspaces of a Hilbert space. 
Also, the set of states is not constrained to the set of pure states.
The responsibility of correct modeling is passed to the specifier.
These ideas led to a proof of initiality which was reported in \cite{gai-qinit}, the contribution where hybrid-dynamic quantum logic was originally defined and the foundation of the present work.
We are not aware of any other Birkhoff completeness result for quantum logics.
Recent studies show that classical computing can aid unreliable quantum processors to solve large problems reliably.
In the future, we are planning to equip our hybrid-dynamic quantum logic with features that support the description of classical programs with quantum subroutines.  
With the development of quantum hardware devices, it becomes increasingly important to develop high-quality and trustworthy quantum software. 
This is possible only by applying formal methods based on solid logical foundations.

\bibliographystyle{aiml}
\bibliography{cut-free}
\appendix
\section{Proofs for results presented in Section~\ref{sec:HDQL}}
\begin{proof}[Lemma~\ref{lemma:inj}: All homomorphisms of Hilbert spaces are injective] \\
Let $h:(W,M)\to(V,N)$ be a homomorphism of quantum models.
For all vectors $w\in W_\vc$, the following are equivalent:

\begin{proofsteps}{15em}
$h(w)=0^V$ &  by inner product definition\\
$\ip{h(w)}{h(w)}^V= 0$ & since $h$ is a first-order homomorphism\\
$h(\ip{w}{w}^W)=0$ & $h$ is the identity on complex numbers\\
$\ip{w}{w}^W=0$ & by the definition of the inner product\\
$w= 0^W$
\end{proofsteps}
Then for all vectors $w_1,w_2\in W_\vc$, 
we have  
$h(w_1)=h(w_2)$ iff 
$h(w_1) -^V h(w_2)=0^V$ iff 
$h(w_1-^W w_2)=0^V$ iff
$w_1-^W w_2=0^W$ iff
$w_1=w_2$.
Hence, $h$ is injective.
\end{proof}

\begin{proof}[Theorem~\ref{prop:sat-cond}: Local satisfaction condition]\\
The proof of the local satisfaction condition is by well-founded (Noetherian) induction on the triple $(n_1,n_2,n_3)$, where 
\begin{enumerate}
\item $n_1$ is the number of occurrences of the operator $*$ in $\gamma$,
\item $n_2$ is the number of occurrences of the operators $\comp$ and $\cup$ in $\gamma$, and
\item $n_3$ is the number of occurrences of the sentence operators in $\gamma$.
\end{enumerate}
We focus only on a few cases, since the rest of them can be discharged using the same arguments from the proof of the satisfaction condition in hybridized institutions~\cite[Theorem 3.2]{DBLP:journals/logcom/Diaconescu16}.
We denote by $(W,M)$ the reduct of $(W',M')$ across $\chi$.
\begin{proofcases}
\item[$p\in\Prop$] 
The following are equivalent:
\begin{proofsteps}{20em}
$(W',M')\models^w \chi(p)$ & by semantics\\
$\chi(p)\in M'_w$ & by the definition of $M_w$ \\
$p \in \{ p\in\Prop \mid \chi(p) \in M'_w \}= M_w$ & by semantics \\
$(W,M)\models^w p$ & 
\end{proofsteps}
\item [$\sim\gamma$] 
By induction hypothesis, we have 
$(W',M')\models^w\chi(\gamma)$ iff $(W,M)\models^w\gamma$ for all vectors $w\in W'_\vc$.
Then:
\begin{proofsteps}{23.5em}
$\chi(\gamma)^{(W',M')}=\gamma^{(W,M)}$ & 
by semantics\\
$(\chi(\gamma)^{(W',M')})^\perp=(\gamma^{(W,M)})^\perp$ & 
as ${W'\!\red_{\Sigma^\hil}}={W\!\red_{\Sigma^\hil}}$\\
$(W',M')\models^w \sim \chi(\gamma)$ iff $(W,M)\models^w\sim \gamma$ 
for all $w\in W'_\vc$ & 
by semantics
\end{proofsteps}
\item[$\nec{f}\gamma$] 
In this case, $f\in U\cup Q$.
The following are equivalent:
\begin{proofsteps}{20em}
$(W',M')\models^w \nec{\chi(f)}\chi(\gamma)$ & by semantics\\
$(W',M')\models^v\chi(\gamma)$, where $v=\chi(f)^{W'}(w)$ & by induction hypothesis\\
$(W',M')\red_\chi\models^v\gamma$ &  
since $v=f^{(W'\red_\chi)}(w)$\\
$(W',M')\red_\chi\models^w\nec{f}\gamma$
\end{proofsteps}
\item[$\nec{\act_1\comp\act_2}\gamma$] \

The number of $*$ in $\nec{\act_1}\nec{\act_2}\gamma$ is equal to the number of $*$ in $\nec{\act_1\comp\act_2}\gamma$.

The number of $\comp$ in $\nec{\act_1}\nec{\act_2}\gamma$ is strictly less than the number of $\comp$ in $\nec{\act_1\comp\act_2}\gamma$.

The number of $\cup$ in $\nec{\act_1}\nec{\act_2}\gamma$ is equal to the number of $\cup$ in $\nec{\act_1\comp\act_2}\gamma$.

By induction hypothesis, we have:
\begin{center}
$(W',M')\models^w\nec{\chi(\act_1)}\nec{\chi(\act_2)}\chi(\gamma)$ iff
$(W',M')\red_\chi\models^w\nec{\act_1}\nec{\act_2}\gamma$.
\end{center}

The following are equivalent:
\begin{proofsteps}{19em}
$(W',M')\models^w\nec{\chi(\act_1)\comp\chi(\act_2)}\chi(\gamma)$ & by semantics\\
$(W',M')\models^w\nec{\chi(\act_1)}\nec{\chi(\act_2)}\chi(\gamma)$ & by induction hypothesis\\ 
$(W',M')\red_\chi\models^w\nec{\act_1}\nec{\act_2}\gamma$ & by semantics \\
$(W',M')\red_\chi\models^w\nec{\act_1\comp \act_2}\gamma$
\end{proofsteps}
\item[$\nec{\act_1\cup\act_2}\gamma$] \

The number of $*$ in $\nec{\act_1}\gamma$ is less or equal to the number of $*$ in $\nec{\act_1\cup\act_2}\gamma$.

The number of $\comp$ in $\nec{\act_1}\gamma$ is less or equal than the number of $\comp$ in $\nec{\act_1\cup\act_2}\gamma$.

The number of $\cup$ in $\nec{\act_1}\gamma$ is strictly less than the number of $\cup$ in $\nec{\act_1\cup\act_2}\gamma$.

By induction hypothesis, we have:
\begin{center}
$(W',M')\models^w\nec{\chi(\act_1)}\chi(\gamma)$ iff ${(W',M')\red_\chi}\models^w\nec{\act_1}\gamma$.
\end{center}

Similarly, induction hypothesis holds for $\nec{\act_2}\gamma$ too.

The following are equivalent:
\begin{proofsteps}{25em}
$(W',M')\models^w\nec{\chi(\act_1)\cup\chi(\act_2)}\chi(\gamma)$ & 
by semantics \\
$(W',M')\models^w\nec{\chi(\act_1)}\chi(\gamma)$ and $(W',M')\models^w\nec{\chi(\act_2)}\chi(\gamma)$ & 
by induction hypothesis\\
${(W',M')\red_\chi}\models^w\nec{\act_1}\gamma$ and ${(W',M')\red_\chi}\models^w\nec{\act_2}\gamma$ & 
by semantics\\
${(W',M')\red_\chi}\models^w\nec{\act_1\cup\act_2}\gamma$
\end{proofsteps}
\item[$\nec{\act^*}\gamma$]
The number of $*$ in $\nec{\act^n}\gamma$ is strictly less than the number of $*$ in $\nec{\act^*}\gamma$,
for all natural numbers $n\in\mathbb{N}$.
By induction hypothesis, we have:
\begin{center}
$(W',M')\models^w\nec{\chi(\act)^n}\chi(\gamma)$ iff 
${(W',M')\red_\chi}\models^w\nec{\act^n}\gamma$ for all $n\in\mathbb{N}$.
\end{center}
The following are equivalent:
\begin{proofsteps}{21em}
$(W',M')\models^w\nec{\chi(\act)^*}\chi(\gamma)$ & 
by semantics\\
$(W',M')\models^w\nec{\chi(\act)^n}\chi(\gamma)$ for all $n\in\mathbb{N}$ &  
by induction hypothesis\\
${(W',M')\red_\chi}\models^w\nec{\act^n}\gamma$ for all $n\in\mathbb{N}$ & 
by semantics \\
${(W',M')\red_\chi}\models^w\nec{\act^*}\gamma$
\end{proofsteps}
\end{proofcases}
\end{proof}
\section{Proofs for results presented in Section~\ref{sec:birkhoff}}
\begin{proof} [Proposition~\ref{prop:basic-comp}: Basic compactness]\\
Let $\vdash$ be the least entailment system closed under the proof rules defined in Fig.~\ref{table:basic}~--~\ref{table:nec}.
Let $\Vdash$ be the compact entailment system defined by $\Gamma\Vdash^k \gamma$ if $\Gamma_f\vdash^k\gamma$ for some finite subset $\Gamma_f\subseteq\Gamma$.
It suffices to show that  $\Vdash$ is an entailment system (i.e., it is closed under $(Monotonicity)$, $(Unions)$ and $(Translation)$) closed under the proof rules defined in Fig.~\ref{table:basic}~--~\ref{table:nec}.
We focus on a few cases that may imply difficulties:
\begin{proofcases}
\item [$(Translation)$]
Assume that $\Gamma\Vdash_{\Delta_\W}^k\gamma$. 
Let $\chi:\Delta_\W\to\Delta'_{\W'}$ be a signature morphism.
By the definition of $\Vdash$, $\Gamma_f\vdash_{\Delta_\W}\gamma$ for some finite $\Gamma_f\subseteq \Gamma$.
Since $\vdash$ is closed under $(Translation)$, $\chi(\Gamma_f)\vdash_{\Delta'_{\W'}}\chi(\gamma)$.
Since $\chi(\Gamma_f)$ is finite, by the definition of $\Vdash$, 
we obtain $\chi(\Gamma)\Vdash_{\Delta'_{\W'}}\chi(\gamma)$.

\item[$(Conj_I)$] 
Assume that $\Gamma\Vdash^k\gamma_1$ and $\Gamma\Vdash^k\gamma_2$.
By the definition of $\Vdash$, 
we have $\Gamma'\vdash^k\gamma_1$ and $\Gamma''\vdash^k\gamma_2$ for some finite subsets $\Gamma'\subseteq \Gamma$ and $\Gamma''\subseteq \Gamma$.
Let $\Gamma_f=\Gamma'\cup\Gamma''$.
By $(Unions)$, $\Gamma_f\vdash^k\gamma_1$ and $\Gamma_f\vdash^k\gamma_2$. 
By $(Conj_I)$, $\Gamma_f\vdash^k\gamma_1\wedge\gamma_2$.
Since $\Gamma_f$ is finite, by the definition of $\Vdash$, we get $\Gamma\Vdash^k\gamma_1\wedge\gamma_2$.
\item[$(Union_I)$] 
Assume that $\Gamma\Vdash^k \nec{\act_1}\gamma$ and  $\Gamma\Vdash^k \nec{\act_2}\gamma$.
By the definition of $\Vdash$, 
we have $\Gamma'\vdash^k  \nec{\act_1}\gamma$ and $\Gamma''\vdash^k  \nec{\act_2}\gamma$ for some finite subsets $\Gamma'\subseteq \Gamma$ and $\Gamma''\subseteq \Gamma$.
Let $\Gamma_f=\Gamma'\cup\Gamma''$.
By $(Unions)$, $\Gamma_f\vdash^k  \nec{\act_1}\gamma$ and $\Gamma_f\vdash^k  \nec{\act_2}\gamma$. 
By $(Union_I)$, $\Gamma_f\vdash^k \nec{\act_1\cup\act_2}\gamma$.
Since $\Gamma_f$ is finite, by the definition of $\Vdash$, we get $\Gamma\Vdash^k \nec{\act_1\cup\act_2}\gamma$. 
\end{proofcases}
\end{proof}

\begin{proof}[Theorem~\ref{th:basic}: Basic completeness]\\
The first statement is proved by well-founded (Noetherian) induction on the triple $(n_1,n_2,n_3)$, where 
\begin{enumerate}
\item $n_1$ is the number of occurrences of the operator $*$ in $\varphi$,
\item $n_2$ is the number of occurrences of the operators $\comp$ and $\cup$ in $\varphi$, and
\item $n_3$ is the number of occurrences of the sentence operators in $\varphi$.
\end{enumerate}

\begin{proofcases}
\item[$p\in\Prop$]
The following are equivalent:
\begin{proofsteps}{20em} 
$\Gamma\vdash^k p$ & by semantics\\
 $p\in M^\Gamma_w$, where $w=k^{(W^\Gamma)}$ & by semantics \\
$(W^\Gamma,M^\Gamma)\models^w p$
\end{proofsteps}
\item[$\at{j}\varphi$]
The following are equivalent:
\begin{proofsteps}{20em} 
$\Gamma\vdash^k \at{j}\varphi$ & 
by $(Ret_I)$ and $(Ret_E)$\\
$\Gamma\vdash^j \varphi$ & 
by induction hypothesis\\
$(W^\Gamma,M^\Gamma)\models^v \varphi$, where $v=j^{(W^\Gamma)}$ & 
by semantics\\
$(W^\Gamma,M^\Gamma)\models^w \at{j}\varphi$, where $w=k^{(W^\Gamma)}$
\end{proofsteps}

\item[$\varphi_1\wedge\varphi_2$] 
The following are equivalent:
\begin{proofsteps}{20em} 
$\Gamma\vdash^k \varphi_1\wedge \varphi_2$ & 
by $(Conj_I)$ and $(Conj_E)$ \\
$\Gamma\vdash^k \varphi_1$ and $\Gamma\vdash^k \varphi_2$ & 
by induction hypothesis\\
$( W^\Gamma,M^\Gamma)\models^w\varphi_1$ and $( W^\Gamma,M^\Gamma)\models^w\varphi_2$, where $w=k^{(W^\Gamma)}$ & 
by semantics\\
$( W^\Gamma,M^\Gamma)\models^w\varphi_1\wedge \varphi_2$
\end{proofsteps}
\item [$\nec{f}\varphi$] In this case $f\in U\cup Q$.
The following are equivalent:
\begin{proofsteps}{18em} 
$\Gamma\vdash^k \nec{f}\varphi$ & 
by $(FT_E)$ and $(FT_I)$ \\
$\Gamma\vdash^{f(k)}\varphi$  & 
by induction hypothesis \\
$(W^\Gamma,M^\Gamma)\models^v\varphi$, 
\newline where $w=k^{(W^\Gamma)}$ and $v=f^{(W^\Gamma)}(w)$ & 
by semantics \\
$(W^\Gamma,M^\Gamma)\models^w\nec{f}\varphi$
\end{proofsteps}
\item [$\nec{\act_1\comp\act_2}\varphi$] \

The number of $*$ in $\nec{\act_1}\nec{\act_2}\varphi$ is equal to the number of $*$ in $\nec{\act_1\comp\act_2}\varphi$.

The number of $\comp$ in $\nec{\act_1}\nec{\act_2}\varphi$ is strictly less than the number of $\comp$ in $\nec{\act_1\comp\act_2}\varphi$.

The number of $\cup$ in $\nec{\act_1}\nec{\act_2}\varphi$ is is equal to the number of $\cup$ in $\nec{\act_1\comp\act_2}\varphi$.

By the induction hypothesis,
for all vector terms $k$, we have:
\begin{center}
$\Gamma\vdash^k \nec{\act_1}\nec{\act_2}\varphi$ iff
$( W^\Gamma,M^\Gamma)\models^w \nec{\act_1}\nec{\act_2}\varphi$, where $w=k^{(W^\Gamma)}$.
\end{center}
The following are equivalent:
\begin{proofsteps}{18em} 
$\Gamma\vdash^k \nec{\act_1\comp\act_2}\varphi$ & 
by $(Comp_I)$ and $Comp_E)$\\
$\Gamma\vdash^k \nec{\act_1}\nec{\act_2}\varphi$ &
by induction hypothesis\\
$(W^\Gamma,M^\Gamma)\models^w\nec{\act_1}\nec{\act_2}\varphi$, 
\newline where $w=k^{(W^\Gamma)}$ & 
by semantics\\
$(W^\Gamma,M^\Gamma)\models^w\nec{\act_1\comp\act_2}\varphi$
\end{proofsteps}
\item [$\nec{\act_1\cup\act_2}\varphi$] \

The number of $*$ in $\nec{\act_1}\varphi$ is less or equal then the number of $*$ in $\nec{\act_1\cup\act_2}\varphi$.

The number of $\comp$ in $\nec{\act_1}\varphi$ is less or equal than the number of $\comp$ in $\nec{\act_1\cup\act_2}\varphi$.

The number of $\cup$ in $\nec{\act_1}\varphi$ is strictly less than the number of $\cup$ in $\nec{\act_1\cup\act_2}\varphi$.

By the induction hypothesis,
for all vector terms $k$, we have:
\begin{center}
$\Gamma\vdash^k \nec{\act_1}\varphi$ iff
$( W^\Gamma,M^\Gamma)\models^w \nec{\act_1}\varphi$, 
where $w=k^{(W^\Gamma)}$.
\end{center}
Similarly, the induction hypothesis holds for $\nec{\act_2}\varphi$ as well.

The following are equivalent:
\begin{proofsteps}{17em}  
$\Gamma\vdash^k \nec{\act_1\cup\act_2}\varphi$ & 
by $(Union_I)$ and $(Union_E)$ \\
$\Gamma\vdash^k \nec{\act_1}\varphi$ and  
$\Gamma\vdash^k \nec{\act_2}\varphi$ & 
by induction hypothesis \\
$(W^\Gamma,M^\Gamma)\models^w\nec{\act_1}\varphi$ and
\newline $(W^\Gamma,M^\Gamma)\models^w\nec{\act_2}\varphi$, 
where $w=k^{(W^\Gamma)}$ & 
by semantics\\
$( W^\Gamma,M^\Gamma)\models^w\nec{\act_1\cup \act_2}\varphi$
\end{proofsteps}
\item [$\nec{\act^*}\varphi$]
The number of $*$ in $\nec{\act^n}\varphi$ is strictly less than the number of $*$ in $\nec{\act^*}\varphi$,
for all natural numbers $n\in\mathbb{N}$.
By the induction hypothesis, for all natural numbers $n\in\mathbb{N}$ and all vector terms $k$, we have:  
\begin{center}
$\Gamma\vdash^k \nec{\act^n}\varphi$ iff $(W^\Gamma,M^\Gamma)\models^w\nec{\act^n}\varphi$, where $w=k^{(W^\Gamma)}$.
\end{center}
The following are equivalent:
\begin{proofsteps}{17em}  
$\Gamma\vdash^k \nec{\act^*}\varphi$ & 
by $(Star_I)$ and $(Star_E)$\\
$\Gamma\vdash^k \nec{\act^n}\varphi$ for all $n\in \N$ & 
by induction hypothesis \\
$(W^\Gamma,M^\Gamma)\models^w\nec{\act^n}\varphi$ for all $n\in\N$, 
\newline where $w=k^{(W^\Gamma)}$ & 
by semantics \\
$(W^\Gamma,M^\Gamma)\models^w\nec{\act^*}\varphi$
\end{proofsteps}
\item[$\store{z}\varphi$] 
The following are equivalent:
\begin{proofsteps}{17em}  
$\Gamma\vdash^k\store{z}\varphi$ &  
by $(Store_I)$ and $(Store_E)$\\
$\Gamma\vdash^k\varphi[z\leftarrow k]$ & 
by induction hypothesis\\
$(W^\Gamma,M^\Gamma)\models^w \varphi[z\leftarrow k]$  
\newline where $w=k^{(W^\Gamma)}$ & 
by semantics\\
$(W^\Gamma,M^\Gamma)\models^w\store{z}\varphi$
\end{proofsteps}
\end{proofcases}
For the second statement, since all sentences in $\Phi$ are basic, by the first statement, we have $(W^\Phi,M^\Phi)\models \Phi$.
\begin{itemize}
\item Since $\vdash^k \subseteq \models^k$, we have that $\Phi\vdash^k\varphi$ implies $\Phi\models^k\varphi$.
\item since $(W^\Phi,M^\Phi)\models \Phi$, we have that $\Phi\models^k\varphi$ implies $(W^\Phi,M^\Phi)\models^{k^{(W^\Phi)}}\varphi$.
\end{itemize}
By the first statement, 
$\Phi\vdash^k\varphi$ iff $(W^\Phi,M^\Phi)\models^{k^{(W^\Phi)}}\varphi$ iff $\Phi\models^k\varphi$. 
\end{proof}
\begin{proof}[Theorem~\ref{th:init}: Initiality] \\
Since the second statement is a direct consequence of the first,
we prove only the first statement by well-founded (Noetherian) induction on $(n_1,n_2,n_3)$, where 
\begin{enumerate}
\item $n_1$ is the number of occurrences of the operator $*$ in $\gamma$,
\item $n_2$ is the number of occurrences of the operators $\comp$ and $\cup$ in $\gamma$, and
\item $n_3$ is the number of occurrences of the sentence operators in $\gamma$.
\end{enumerate}
We focus only on implication since the remaining cases can be discharged using arguments from the proof of Theorem~\ref{th:basic}.
\begin{proofcases}
\item [$\varphi\Rightarrow \gamma$]
Assume that $\Gamma\vdash^k \varphi\Rightarrow\gamma$,
where $\varphi$ is basic and $\gamma$ is a quantum clause.

\begin{proofsteps}{15em}
assume that $(W^\Gamma,M^\Gamma)\models^{k^{(W^\Gamma)}} \varphi$ &  \\
$\Gamma\vdash^k \varphi$ &
by Theorem~\ref{th:basic},
since $\varphi$ is basic\\
$\Gamma\vdash^k \gamma$ & 
by $(MP)$, as $\Gamma\vdash^k \varphi\Rightarrow\gamma$ and $\Gamma\vdash^k \varphi$ \\ 
$(W^\Gamma,M^\Gamma)\models^{k^{(W^\Gamma)}} \gamma$ & 
by induction hypothesis
\end{proofsteps}
It follows that $(W^\Gamma,M^\Gamma)\models^{k^{(W^\Gamma)}} \varphi\Rightarrow \gamma$.
\item[$\rho_1\leadsto\rho_2$]
Assume that $\Gamma\vdash^k\rho_1\leadsto\rho_2$, where $\rho_1$ is a closed basic sentence and  $\rho_2$ is a closed quantum clause.
\begin{proofsteps}{14em}
assume $(W^\Gamma,M^\Gamma)\models^{k^{(W^\Gamma)}}\rho_1$ &\\
$\Gamma\vdash^k\rho_1$ & 
by Theorem~\ref{th:basic}, since $\rho_1$ is basic\\
$\Gamma\vdash^k\rho_2$ & 
by $(MP_t)$, as $\Gamma\vdash^k\rho_1\leadsto\rho_2$ and $\Gamma\vdash^k\rho_1$\\
$(W^\Gamma,M^\Gamma)\models^w\rho_2$ & 
by induction hypothesis
\end{proofsteps}
It follows that $(W^\Gamma,M^\Gamma)\models^w \rho_1\leadsto \rho_2$.
\end{proofcases}
\end{proof}
\begin{proof}[Theorem~\ref{th:birkhoff}: Birkhoff completeness]\\
We prove the forward implication (completeness) by well-founded (Noetherian) induction on $(n_1,n_2,n_3)$, where 
\begin{enumerate}
\item $n_1$ is the number of occurrences of the operator $*$ in $\gamma$,
\item $n_2$ is the number of occurrences of the operators $\comp$ and $\cup$ in $\gamma$, and
\item $n_3$ is the number of occurrences of the sentence operators in $\gamma$.
\end{enumerate}
We focus only on implication, since the remaining cases can be discharged using arguments from the proof of Theorem~\ref{th:basic}.
\begin{proofcases}
\item[$\varphi\Rightarrow \gamma$]
$\Gamma\models^k\varphi\Rightarrow \gamma$, where $\varphi$ is a basic sentence and $\gamma$ is a quantum clause, iff 
$\Gamma\cup\{\at{k}\varphi\}\models^k\gamma$ iff (by induction hypothesis)
$\Gamma\cup\{\at{k}\varphi\}\vdash^k\gamma$. 
By $(Imp)$, $\Gamma\vdash^k\varphi\Rightarrow \gamma$.
\item[$\rho_1\leadsto \rho_2$]
$\Gamma\models^k\rho_1\leadsto \rho_2$, 
where $\rho_1$ is closed basic sentence and $\rho_2$ is a closed quantum clause, iff 
$\Gamma\cup\{\at{k}\rho_1\}\models^k\rho_2$ iff (by induction hypothesis)
$\Gamma\cup\{\at{k}\rho_1\}\vdash^k\rho_2$. 
By $(Imp_c)$, $\Gamma\vdash^k\rho_1\leadsto \rho_2$.
\end{proofcases}
\end{proof}

\end{document}